\documentclass[]{elsarticle}

\usepackage{lineno}
\usepackage{amsmath}
\usepackage[psamsfonts]{amssymb}
\usepackage[unicode, pdftex]{hyperref}
\usepackage{color}
\usepackage[margin=2.0cm]{geometry}

\modulolinenumbers[5]

\journal{Journal of \LaTeX\ Templates}

\begin{document}

\begin{frontmatter}

\title{Determination of quantum yield of NADH and FAD in alcohol-water solutions: the analysis of radiative and nonradiative relaxation pathways.}


\author[mymainaddress]{Ioanna A. Gorbunova}

\author[mymainaddress]{Marina K. Danilova}

\author[mymainaddress]{Maxim E. Sasin}

\author[mymainaddress]{Victor P. Belik}

\author[mymainaddress,mysecondaryaddress]{Dmitrii P. Golyshev}

\author[mymainaddress]{Oleg S. Vasyutinskii\corref{mycorrespondingauthor}}
\cortext[mycorrespondingauthor]{Corresponding author}
\ead{osv@pms.ioffe.ru}

\address[mymainaddress]{Ioffe Institute, 26 Polytekhnicheskaya, St.Petersburg, 194021 Russia}
\address[mysecondaryaddress]{Peter the Great St.Petersburg Polytechnic University, 29 Polytechnicheskaya, St.Petersburg, 195251, Russia}

\begin{abstract}
 Combined studies on fluorescence quantum yield in coenzymes NADH and FAD in water-methanol, water-ethanol and water-propylene glycol  mixtures and on time-resolved fluorescence of the same molecules by means of the TCSPC method under excitation at 450 and 355~nm have been carried out. A significant difference in the behavior of NADH and FAD quantum yields in different alcohol-water mixtures was observed. The dependence of quantum yield in NADH on alcohol concentration was found to be similar in monohydric alcohols methanol and ethanol and differed from that in polyhydric alcohol propylene glycol. In the case of FAD, the behavior of quantum yield was almost independent of the type of alcohol and exhibited a dramatic increase of 5--6 times with alcohol concentration. The experimental results were analysed using a model based on the quantum mechanical theory developed recently by the authors that took into account radiative and nonradiative relaxation pathways.  A generalised expression for the fluorescence quantum yield containing contributions from picosecond and nanosecond nonratiative decay processes in molecular excited states was derived and used for clarification of the role of nanosecond and picosecond relaxation channels in NADH and FAD in various water-alcohol mixtures. The analysis of theoretical expression describing  fluorescence decay in polyatomic molecules excited by a short laser pulse resulted in a new insight into the nature of the Decay Associated Spectra (DAS) phenomenon.
The contribution from the fast picosecond nonradiative decay in NADH to the quantum yield in water-methanol and water-ethanol mixtures was found to be practically independent of the alcohol concentration and hence of the NADH conformation distribution. This result suggests that the picosecond decay in NADH does not likely occur through electron transfer in the stacking configuration of the nicotinamide and adenine moieties but through other mechanisms. At the same time the contribution of the fast picosecond nonradiative decay in NADH to the quantum yield in water-propylene glycol solutions was found to depend on propylene glycol concentration, therefore the sharp nonlinear increase of the measured quantum yield was associated with decrease of both fast picosecond and relatively slow nanosecong quenching nonradiative rates. In FAD, the contribution of the fast picosecond quenching to the quantum yield was found to depend significantly on the alcohol concentration in all alcoholes under study. This finding suggests that the fast quenching gives a profound contribution to the rise of the measured quantum yield with alcohol concentration and supports the established mechanism of the fluorescence quenching in FAD through electron transfer reaction in the $\pi$-stacked conformation between isoalloxazine and adenine moieties.
\end{abstract}

\begin{keyword}
fluorescence quantum yield, NADH, FAD, solution, fluorescence quenching, fluorescence decay times, picosecond relaxation, multiexponential decay, alcohol-water mixture
\end{keyword}

\end{frontmatter}

\section{Introduction}
\label{sec:Introduction}
Reduced nicotinamide adenine dinucleotide (phosphate) (NAD(P)H) and oxidized flavin adenine dinucleotide (FAD) are essential biological coenzymes involved in regulation of living cell metabolism that are actively used  nowadays as intrinsic fluorescent biomarkers for monitoring the respiratory chain activity and cell metabolism~\cite{Heikal2010,Shirmanova2018,Ma2016}.  A high potential of NAD(P)H and FAD autofluorescence as biomarkers was highlighted in a series of seminal works by Chance et al.~\cite{Chance1962,Chance1965}.  Chance et al. have demonstrated that NADH  in its reduced form exhibits autofluorescence at 400--500~nm, whereas its oxidized form, NAD$^+$ does not fluoresce and developed spectroscopic ways to measure a redox ratio  NAD$^+$/NADH for real-time monitoring of the metabolic state of a cell during pathophysiological changes.

Reduced FAD (FADH$_2$) and FAD are another redox pair largely associated with respiration in mitochondria of eukaryotic cells. As known, the reduced form FADH$_2$ does not fluoresce while the oxidized form FAD can fluoresce under excitation in the spectral region shorter than 420~nm~\cite{Islam2003}. Zhang et al.~\cite{Zhang2004} suggested a prominent way of determination the mitochondria redox state by measuring the redox ratio HADH/FAD and shown that the measurement of the ratio of fluorescence has important advantages over the absolute values.

Chance et al. \cite{Chance1962, Barlow1976} suggested that the changes occurring in redox state in cancer cells can be effectively studied by monitoring NADH fluorescence. This approach was developed and widely used
by many groups (see, e.g. \cite{Kasischke2004,Blinova2005}) however further experiments elucidated a number of biological and technical problems dealing with light scattering in tissue, absorption of the laser light at
360~nm by other intracellular species, and the influence of a number of factors on fluorescence intensity.

To overcome these problems Lakowicz et al.~\cite{Lakowicz1992} employed the measurements of NADH fluorescence lifetimes that are independent of light intensity and can be used for separation of free and protein-bound NADH quantitatively. Moreover, Bird et al.~\cite{Bird2005} reported that the ratio of free/protein-bound NADH is related to changes in the NADH/NAD$^+$ ratio. Therefore, quantifying NAD(P)H and FAD fluorescence through fluorescence lifetime imaging  allows in general to determine the optical redox ratio with high sensitivity and therefore to monitor small changes in metabolic activity~\cite{Ma2016}. Fluorescence lifetime imaging microscopy (FLIM) is now widely used for monitoring of metabolic pathways activated in cancer cells~\cite{Shirmanova2018,Yaseen2017,Evers2018,Schaefer2019}.

As demonstrated in numerous studies mentioned above, FLIM in NAD(P)H and FAD can be used for non-destructive determination of changes in cellular metabolism. However, till now these changes are difficult to interpret \cite{Sharick2018}. The isotropic (polarization insensitive) part of the fluorescence signal $I_{tot}(t)$ is usually presented as a sum of exponentials~\cite{Blacker2013,Hull2001,Blacker2019,Gorbunova20a,Gorbunova21,vandenberg2002,Chosrowjan2003,Kao2008,Radoszkowicz2011,Knutson2019,Knutson2020}:
\begin{equation}
\label{sum}
 I_{tot}(t)= I_0\sum\limits_{i=1}^na_{i}\,\exp\left(-\frac{t}{\tau_{i}}\right),
\end{equation}
where $I_0$ is a time-independent initial fluorescence intensity, $\tau_i$ are decay times, and $a_i$ are corresponding weighting coefficients that are normalized to unity: $\sum_i a_i=1$.

Nowadays using several supplementary fluorescence detection experimentalist can measure a great number of fluorescence decay times $\tau_i$ describing various aspects of the fluorescence dynamics. One of them is time correlated single photon counting (TCSPC) technique allowing for determination of decay times in the nanosecond and near sub-nanosecond time domain. The others are femtosecond transient monitoring and up-conversion techniques that allow for determination of the decay times in picosecond and sub-picosecond time domain.

In the nanosecond and near sub-nanosecond time domain free NADH in aqueous solution is known to exhibit biexponential fluorescence decay with the lifetimes of about 0.3~ns and 0.7~ns~\cite{Visser1981,Couprie1994a,Hull2001,Vasyutinskii2017} and in other solvents the lifetimes and corresponding weighting coefficients depend strongly on temperature and solvent type \cite{Ladokhin1995, Couprie1994a, Visser1981,Blacker2013}. The explanation of the nature of the two observed fluorescence decay times in NADH for long time remained controversial. In particular it was unclear if the two fluorescence lifetimes either correspond to different geometric conformations of the whole molecule, or they are intrinsic features of the NA ring~\cite{Visser1981,Gafni1976,Ladokhin1995,Krishnamoorthy1987,Kierdaszuk1996,Blacker2013,Blacker2019}. Gorbunova et al.~\cite{Gorbunova20c} suggested that the heterogeneity in the measured decay times in NADH is due to different charge distributions in the \emph{cis} and \emph{trans} configurations of the nicotinamide ring that results in different electrostatic field distributions and different non-radiative decay rates $(\tau^{\,nrad}_l)^{-1}$ (see eq.~(\ref{eq:gammai})). This model was also shown to be useful for elucidation of the single-exponential and relatively slow fluorescence decay in NADH-alcoholedegidrogenase (ADH) complex~\cite{Gorbunova21} because  as known from the X-ray structure spectroscopy and NMR experiments, NAD(P)H imbedded into the ADH binding site in a single unfolded \emph{anti} "\emph{trans}-out-of-plane" nuclear conformation~\cite{Hammen2002,Plapp17,Vidal2018}. Within the same model a significant enhancement of the decay time value in the ADH-bounded NADH compare with the free NADH in solution can be attributed~\cite{Gorbunova21} to a significant decrease of charges separations in the nicotinamide (NA) ring in the conditions of a typical apolar binding site environment in the ADH-NADH complex~\cite{Piersma1998,Vishwasrao2005}.  Similar reasons are likely responsible for significantly different lifetime values observed in NADH bounded with various enzymes \cite{Sharick2018}.

In the nanosecond time domain FAD in solutions and living cells demonstrates a multiexponential fluorescence decay dynamics.  At physiological pH and temperature conditions in solution it is most common to fit FAD fluorescence decay after excitation of the isoalloxazine moiety by two exponentials with the lifetimes of about 2.3~ns and 3.5 to 4.5~ns~\cite{Sengupta2014,Nakabatashi2010,Islam2013,Krasnopevtseva2020}. However, in other solvents the number of lifetimes, their values and corresponding weighting coefficients depend strongly on temperature, pH and solvent type~\cite{Drossler2002,Islam2003,vandenberg2002,Nakabatashi2010,Sengupta2014}. Despite numerous studies carried out the origin of the two fluorescence nanosecond decay times in FAD still remains poorly understood.

Multiexponential fluorescence decay was also observed in the picosecond time domain by many groups using ultrafast fluorescence up-conversion and transient monitoring methods.  A very short decay time of about $\tau \simeq$ 1~ps that was observed by several groups in NADH~\cite{Knutson2019,Knutson2020,Peon2020}, FAD~\cite{Radoszkowicz2011,Nakabatashi2010,vandenberg2002,Chosrowjan2003,Kondo2006,Kao2008}, and many other molecules (see, e.g.~\cite{Fleming94,Knutson2009,Xu2008}). It is commonly attributed to water relaxation around the corresponding excited chromophore moiety. A signature of this vibrational relaxation is Decay Associated Spectra (DAS) that is characterised by the positive value of the corresponding pre-exponential weighting coefficient $a_i$ in eq.~(\ref{sum}) in the blue wing of the fluorescence band and negative, or near zero coefficient  $a_i$ in the red wing of the fluorescence band~\cite{Knutson2019,Peon2020,Kao2008,Xu2008}. However, the understanding of DAS still remained controversial~\cite{Knutson2009}.

Another typical picosecond decay times of about $\tau \simeq$ 26~ps observed in NADH~\cite{Knutson2019,Knutson2020} and of about $\tau_d \simeq$ 5-9~ps observed in FAD~\cite{Chosrowjan2003,Kao2008,Nakabatashi2010,Radoszkowicz2011} were associated with picosecond relaxation occurring between the excited $S_1$ and the ground $S_0$ electronic states and resulting in fast fluorescence quenching. These fluorescence decay is known as quasi-static-self-quenching (QSSQ)~\cite{Xu2008} and results in decrease of the observed quantum yield. In the case of NADH the routes of QSSQ are still not known exactly~\cite{Knutson2019,Knutson2020,Peon2020}. In particular, these can be either interactions of excited NADH molecules with surrounding solvent molecules, or intramolecular internal conversion to the ground electronic state, or intersystem crossing to the triplet manifold~\cite{Lakowicz97a}.
In the case of FAD it is now well adopted that the mechanism of the fluorescence quenching is an electron transfer reaction in the $\pi$-stacked conformation~\cite{Islam2003,Li2008,Navarro2016,vandenberg2002,Radoszkowicz2010}.

A common way for determination of the fractional population of free and bound species from the multiexponential expansion in eq.~(\ref{sum}) is to assume that the radiation decay rate $\gamma_{rad}=\tau_{rad}^{-1}$ remains constant and then suggest that the weighting coefficients $a_i$ in eq.~(\ref{sum}) are equal to the fractional concentrations of the species associated with it and to use the fraction of the total fluorescence signal generated by each species $i$~\cite{Vishwasrao2005,Niesner2004,Sharick2018}:
\begin{equation}
\label{f}
f=\frac{a_i\tau_i}{\sum_j a_j\tau_j }.
\end{equation}

However, the origin of the fluorescence decay times numbers and their values in NADH and FAD bound to different enzymes still remains controversial. Therefore, the relative number of coenzyme molecules bound to different enzymes cannot be obtained without further characterization~\cite{Niesner2004,Cao2020,Gorbunova21}. As pointed out by Niesner et al.~\cite{Niesner2004} in general $a_i$ and $f$ values are influenced by the concentration, absorption cross-section, and fluorescence quantum yield. Consequently, it is necessary to know the photophysical properties of the fluorescing species in order to calculate their relative concentration.

Fluorescence quantum yield is one of the most important parameter needed for determination of the redox ratio and the ratio of free and protein-bound NADH in biological solutions and living cells from fluorescence lifetime measurement \cite{Ma2016,Cong2019} as it is known to be highly dependent on solution/intracellular viscosity and polarity~\cite{Freed1967,Scott1970,Gafni1976,Visser1981,Fischer1988,Couprie1994,Blacker2013}.  Scott et al.~\cite{Scott1970} studied fluorescence lifetime and  quantum yield  of NADH and its derivatives in water and propylene glycol (PG) and revealed that the excited state dynamics in NADH are mostly characterized by non-radiative processes. This suggestion was supported in the following studies~\cite{Gafni1976, Blacker2013, Visser1981,Couprie1994, Fischer1988} where the dependence of fluorescence quantum yield and non-radiative decay rates in NADH on temperature or solvent type was extensively studied. Gafni et al \cite{Gafni1976}, Fischer et al.~\cite{Fischer1988}, and Visser et al.~\cite{Visser1981} reported an increase in the NADH quantum yield by an order of magnitude upon binding with enzymes. Nakabayashi et al.~\cite{Nakabayashi2014} studied the effect of the local electric field on NADH fluorescence in solutions and demonstrated an increase in average fluorescence decay time due to the decrease in solvent polarity.  Blacker et al.~\cite{Blacker2013} inquired  time-resolved fluorescence in NADH in water-glycerol solution and calculated  non-radiative decay rate using Kramers-Hubbard equation.

However, the origin of non-radiative pathways accounting for excited-state dynamic in FAD and especially NADH still remain controversial.  Also it is unclear ether intermolecular quenching processes  or intramolecular interaction of nicotinamide and adenine can be account for the very low fluorescence quantum yield in NADH~\cite{Blacker2013,Scott1970,Fischer1988}.



Fluorescence quantum yield in FAD in solutions and proteins was intensively studied during latest decades~\cite{Islam2003,Nakabatashi2010,Radoszkowicz2010,Sengupta2011}. Nakabayashi et al.~\cite{Nakabatashi2010} observed rapid growth of fluorescence excited at 410 nm depending on alcohol concentration increase in alcohol-water solutions and suggested that the addition of a less polar solvent breaks the $\pi$-$\pi$ interaction in the stacked conformation and produces the open FAD conformation which results in the rise of fluorescence intensity. Radoszkowicz et al.~\cite{Radoszkowicz2010} performed molecular dynamics simulations of FAD in water-methanol mixtures and showed that methanol affected the dynamics of FAD by enhancing the frequency of unfolding events without any effect on the lifetime of unfolded states.

Despite a number of studies carried out the detailed physical interpretation of experimental data in the form of multiexponential expansion in eq.~(\ref{sum}) and in the form of fluorescence quantum yields is still a challenging problem.

This paper aims to address this problem, at least partly. We carried combined experimental studies including the determination of fluorescence quantum yields in NADH and FAD dissolved in water mixtures with several alcohols: methanol (MeOH), ethanol (EtOH), and propylen glycol (PG), and the time-resolved fluorescence studies of the same molecules using TCSPC method under excitation at 450 and 355~nm. The experimental results were analysed using a new model taking into consideration radiative excitation and fluorescence channels as well as nonradiative excited state relaxation pathways and based on the quantum mechanical theory developed recently by the authors~\cite{Gorbunova20b}.

The main results obtained are as follows. A significant difference was observed in the behavior of NADH and FAD quantum yields in alcohol solution. In particular, the dependence of quantum yield in NADH on alcohol concentration was similar for monohydric alcohols (methanol and ethanol) and significantly different for polyhydric alcohol (propylene-glycol). In the case of FAD, the behavior of quantum yield was almost independent of the type of alcohol and exhibited a dramatic increase of 5--6 times with alcohol concentration.

The analysis of general theoretical expression describing  fluorescence decay in a polyatomic molecule excited by a short laser pulse led to better understanding of the nature of the DAS phenomenon. A generalised expression for the fluorescence quantum yield that allowed to separate contributions from the fast and slow nonratiative processes in molecular excited states was derived  and used for clarification of the role of nanosecond and picosecond relaxation channels in NADH and FAD in various water-alcohol mixtures.

The contribution of the fast picosecond nonradiative decay in NADH to the quantum yield in water-MeOH and water EtOH mixtures was found to be practically independent of the alcohol concentration and hence of the NADH conformation distribution. This result suggests that the picosecond decay in NADH does not likely occur through electron transfer in the stacking configuration of the nicotinamide and adenine moieties but through other mechanisms. It also suggests that the known growth of NADH quantum yield at high alcohol concentration occurs mostly due to relatively slow nanosecond nonradiative mechanisms related with conformation distributions.  At the same time the contribution of the fast picosecond nonradiative decay in NADH to the quantum yield in water-PG solutions was found to depend on PG concentration and the sharp nonlinear increase of the measured quantum yield with PG concentration was associated with decrease of both fast picosecond and relatively slow nanosecong quenching nonradiative rates.

In FAD, the contribution of the fast picosecond quenching to the quantum yield was found to depend significantly on the alcohol concentration in all MeOH, EtOH. This finding suggests that the fast quenching gives a profound contribution to the rise of the measured quantum yield with alcohol concentration and  supports the established mechanism of the fluorescence quenching in FAD through electron transfer reaction in the $\pi$-stacked conformation between isoalloxazine and adenine moieties.


\section{Experimental Procedure}
\subsection{Materials}
FAD disodium salt (98\% purity, DIA-M) and $\beta$ -- NADH (NADH disodium salt, 98 \% purity, Sigma-Aldrich) were used in experiments. Unbuffered distilled water (pH = 5.5),  methanol (MERCK, extra pure), ethanol (extra pure), propylene glycole and their mixtures with water were used as solvents.  The 1.5 ml water-alcohol mixtures of various concentrations were preliminary prepared and then 25$\mu$l FAD or NADH stock solutions were added to the water-alcohol mixtures. Final FAD and NADH concentrations were equal to 60 $\mu$M and 15 $\mu$M respectively.  These concentrations was chosen since similar FAD and NADH concentration were used in earlier studies~\cite{Sengupta2014, Islam2003, Krasnopevtseva2020, Scott1970} that allowed to compare our experimental results with others in similar conditions.  FAD and NADH concentration used in the experiments was much less than its solubility,therefore intramolecular interaction between dissolved molecules and aggregate formation could be neglected.  All solutions were prepared fresh daily and used at the temperature of 20$^\circ$C.

\subsection{Quantum yield measurements}
The procedure of the absolute fluorescence quantum yield determination in NADH and FAD molecules in water-alcohol solutions was similar to that used in previous studies (see, e.g.~\cite{Nakabatashi2010,Radoszkowicz2010, Fischer1988, Scott1970}). In brief, the experimental procedure was as follows.  An excitation laser beam propagated through a quartz cuvette (10x10 mm) containing the molecular solution. All experiments were carried out in the low absorption condition when the absorption in the experimental sample was less than 15\% to avoid saturation and photobleaching effects. The  laser beam  intensity transmitted through the cuvette with the molecular solution $\langle I_{tr}^{mol}\rangle$ and with pure solvent $\langle I_{tr}^{sol}\rangle$ was recorded by a phododiode to determine absorption in the molecular sample. Molecular fluorescence was collected at the right angle to the laser beam propagation, then it passed through the glass optical filter for damping the scattered excitation light, and was focused onto a photodetector. The absolute quantum yield was determined using the expression:
\begin{equation}
\label{eq:AbsQ}
    Q = NG \frac{\langle I_{fl}\rangle}{\langle I^{sol}_{tr}\rangle - \langle I^{mol}_{tr}\rangle},
\end{equation}
where $\langle I_{fl}\rangle$ is the fluorescence intensity, angle brackets mean time averaging, $N$ is a constant taking into account the photon energy of the excitation and fluorescence light, and G is the geometry factor calculated as the ratio of the solid angle within fluorescence detection area to the total emission solid angle.

As known, FAD has two low laying wide absorption bands with maxima at 450 and 370~nm and NADH has a wide absorption band with a maximum at 340 nm. A pulsed  Nd:YAG laser (SNV-40P, Teem Photonics) with output at 355 nm, pulse duration of 400~ps, and repetition rate of 2~KHz was used in our experiments for excitation of NADH and  for excitation of FAD through the second absorption band, and a pulsed semiconductor laser with output at 450 nm, pulse duration of 30~ps, and repetition rate of 20~MHz  was used for excitation of FAD through the first absorption band. Only the time-averaged laser intensities were recorded by the photodetector. Each laser intensity at the cuvette was kept at about 40 mW. The fluorescence intensities were recorded in NADH and FAD within whole their fluorescence bands without spectral resolution.  The molecular fluorescence and laser beam intensities transmitted through the cuvette were detected by a photodiode photodetector PD300-SH (Ophir Optronics) with the sensitivity spectral range of 350--1100~nm. The photodetector was calibrated taking into account its wavelengths sensitivity.

In most of the experiments for increasing of the signal-to-noise ratio the laser beam was modulated at the frequency of 114 Hz by a chopper installed in front of the cuvette. An amplitude-modulated electric signal from the photodetector was separated and detected by a narrow-band lock-in amplifier and then collected and analysed with a computer. An intensive fluorescence signal in FAD in aqueous solution was recorded in a wide-band frequency regime without laser beam modulation and then used for determination of the absolute quantum yield with respect to eq.~(\ref{eq:AbsQ}). This signal was used for calibration of all other signals. Having in mind that the time-averaged fluorescence polarization was  relatively low  in all our experiments (less than 14\% in NADH and less than 4\% in FAD) no special corrections of the fluorescence amplitude due to polarization effects were made as these corrections were smaller that the quantum yield experimental error bars shown in Figs.~\ref{fig:qy_FAD__alc} and \ref{fig:qy_NADH_alc} below.

\subsection{Lifetimes measurements}
The procedure of determination of the fluorescence decay times was similar to that described in detail in our recent publications~\cite{Denicke10,Herbrich15,Sasin18,Gorbunova20c,Gorbunova21}. Briefly, the pulsed lasers with output at 450~nm and 355~nm described above were used in this paper for one-photon excitation of FAD via the first and the second absorption bands. Two orthogonally polarized fluorescence components $I_{\|}(t)$ and $I_{\bot}(t)$ were recorded in experiment using time correlated single photon counting (TCSPC) technique. The polarization components $I_{\|}(t)$ and $I_{\bot}(t)$ was fitted with the multiexponential function in eq.~(\ref{sum}) using a global fit procedure with a least square differential evolution algorithm with convolution~\cite{Gorbunova20c}. The polarization-insensitive part of the fluorescence intensity $I_{tot}(t)$  was calculated according to:
\begin{equation}
\label{eq:Itot}
 I_{tot}(t)=I_{\|}(t)+2I_{\bot}(t).
\end{equation}

\section{Experimental results}
\label{sec:experimental results}

\begin{figure} [ht]
   \begin{center}
   \begin{tabular}{c}
   \includegraphics[height=5cm]{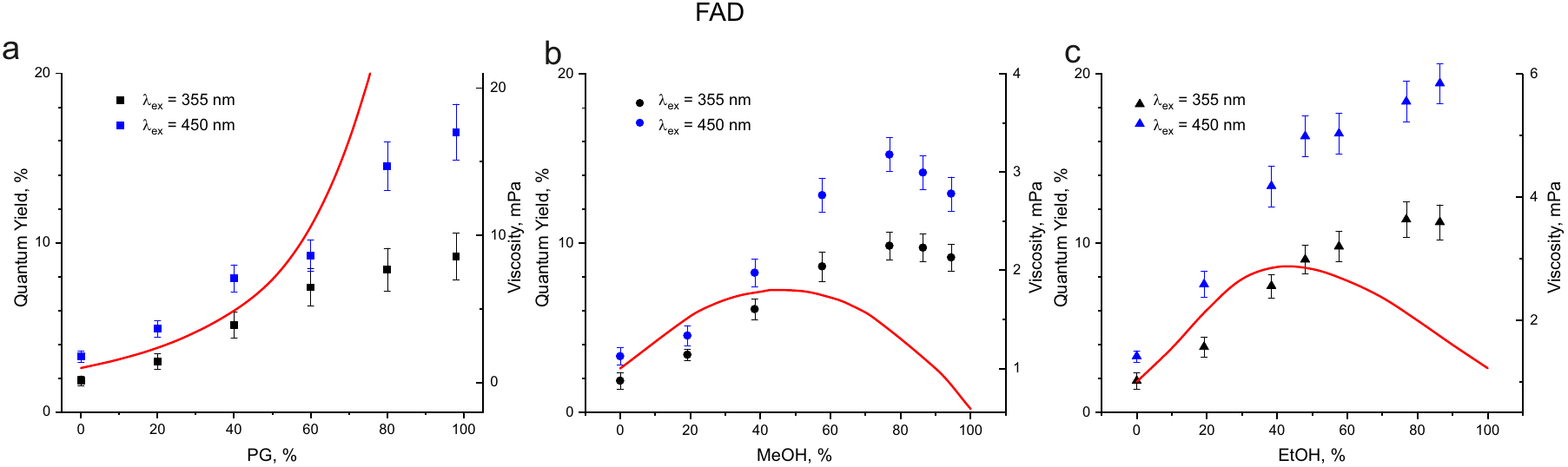}
   \end{tabular}
   \end{center}
   \caption[]{Fluorescence quantum yield in FAD upon excitation at 450 nm and 355 nm in three water-alcohol solutions: propylene glycole (a), methanol (b), and ethanol (c) as a function of alcohol concentration. Symbols in plots a--c are experimental data and red curves in represent the corresponding alcohol viscosities.}
    \label{fig:qy_FAD__alc}
\end{figure}

Fluorescence quantum yield in FAD determined upon excitation at 450 nm and 355 nm in three different water-alcohol solutions of various concentrations are presented in  Figs.~\ref{fig:qy_FAD__alc}a--c as a function of the alcohol concentration.

According to Figs.~\ref{fig:qy_FAD__alc}a--c the absolute quantum yield in FAD in pure water under excitation at 450 nm and 355 nm was found to be  $Q$=3.33$\pm$\% and $Q$=1.87$\pm$\%, respectively. The quantum yield in pure water at 450 nm in Figs.~\ref{fig:qy_FAD__alc}a--c agrees perfectly with that reported earlier by Islam et al.~\cite{Islam2003} and Sengupta et al.~\cite{Sengupta2011}. To the best of our knowledge the quantum yield in FAD under excitation at 355 nm was determined in this paper for the first time.

As can be seen in Fig.~\ref{fig:qy_FAD__alc} the fluorescence quantum yield in FAD rose dramatically with alcohol concentration in all solutions under study at each excitation wavelength. In water-propylene glycol solutions fluorescence quantum yield in Fig.~\ref{fig:qy_FAD__alc}a increased practically linearly with propylene glycol concentration and in pure propylene glycol reached the values of $Q$=17\% and $Q$=9\% under excitation at 450 and 355~nm, respectively. In water-methanol solutions  the quantum yield in Fig.~\ref{fig:qy_FAD__alc}b also increased linearly up to the methanol concentration of about 80\% and reached the values of $Q$=15\% and $Q$=10\% under excitation at 450 and 355~nm, respectively. At higher methanol concentrations the quantum yield decreased somehow by 2--3\% at both excitation wavelengths. In water-ethanol solutions  the quantum yield in Fig.~\ref{fig:qy_FAD__alc}c increased almost linearly with alcohol concentration up to the concentration of about 70\% and then increased more slow at higher concentrations reaching the maxima of 19\% and 11\% under excitation at   450 and 355~nm, respectively.

\begin{figure} [ht]
   \begin{center}
   \begin{tabular}{l}
   \includegraphics[height=5cm]{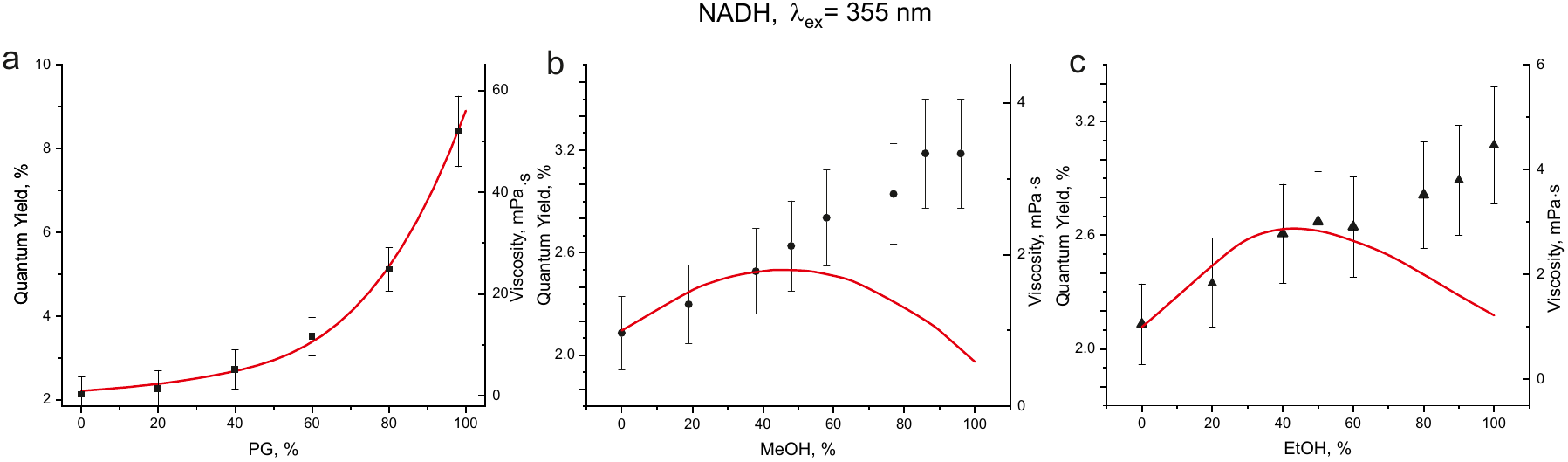}
   \end{tabular}
   \end{center}
   \caption{Fluorescence quantum yield in NADH upon excitation at 355 nm in three water-alcohol solutions: propylene glycole (a), methanol (b), and ethanol (c) as a function of alcohol concentration. Black symbols in plots  a--c are experimental data and red curves in represent the corresponding alcohol viscosities.}
    \label{fig:qy_NADH_alc}
\end{figure}

Quantum yields in NADH determined upon excitation at 355 nm in water--propylene glycol, water-methanol, and water-ethanol solutions as a function of alcohol concentrations are shown in  Figs.~\ref{fig:qy_NADH_alc}a--c.

The fluorescence quantum yield in aqueous solution was determined to be $Q$=2.1$\pm$0.2 that agrees perfectly with the value reported earlier by Scott et al.~\cite{Scott1970} and Gafni et al.~\cite{Gafni1976}.  As can be seen in Fig. \ref{fig:qy_NADH_alc}a in water-propylene glycol solutions, fluorescence quantum yield in NADH  exhibited a nonlinear behavior with propylene glycol concentration and reached the maximum of $Q$=8\% in pure propylene glycol. This result is also in agreement with the earlier result reported by Scott et al.~\cite{Scott1970}. The quantum yield dependence in Fig.~\ref{fig:qy_NADH_alc}a almost perfectly follows the propylene glycol viscosity dependence, also the quantum yield value in pure propylene glycol was about 4 times larger than in water.  As can be seen in Figs.~\ref{fig:qy_NADH_alc}b,c in the case of water-methanol and water-ethanol solutions fluorescence quantum yield in NADH increased practically linearly with alcohol concentration and reached the maxima of about 3.2 \% in the pure alcoholes. As can be seen in Figs.~\ref{fig:qy_NADH_alc}b,c fluorescence quantum yield in NADH was about 1.5 times larger in pure alcohol than in water.

The dependence of FAD and NADH quantum yields  on ethanol and methanol concentrations  was studied earlier by several groups~\cite{Nakabatashi2010,Radoszkowicz2010,Fischer1988, Scott1970}. Our results presented in Figs.~\ref{fig:qy_FAD__alc} and \ref{fig:qy_NADH_alc} are at least in qualitative  agreement with them.  The increase of fluorescence intensity in FAD in water-ethanol and water-glycerol solutions was observed early by Nakabayashi et al~\cite{Nakabatashi2010} who observed an almost 5 times increase of FAD fluorescence intensity upon excitation at 445 nm in 50\% ethanol solution compared with pure water.  The dependence of FAD quantum yield on methanol concentration shown in Fig.~\ref{fig:qy_FAD__alc}b is consistent with the result reported earlier by Radoszkowicz et al \cite{Radoszkowicz2010}. In the case of NADH, an 1.5 times increase of fluorescence intensity upon excitation at 360 nm in methanol solution compared with water was reported by Freed et al.~\cite{Freed1967}.  Moreover, the same increase in fluorescence intensity and relative quantum yield in dihydronicotinamide in ethanol and methanol was observed by Fischer et al \cite{Fischer1988}. At the same time, the complete quantitative data on fluorescence quantum yields in NADH and FAD in Figs.~\ref{fig:qy_FAD__alc} and \ref{fig:qy_NADH_alc} are presented in this paper at the first time.

The experimental results shown in Figs.~\ref{fig:qy_FAD__alc} and \ref{fig:qy_NADH_alc} can be briefly summarized as follows. A significant difference was observed in the behavior of NADH and FAD quantum yields in alcohol solution. In particular, the dependence of quantum yield in NADH on alcohol concentration was similar for monohydric alcohols (methanol and ethanol) and significantly different for polyhydric alcohol (propylene-glycol). In the case of FAD, the behavior of quantum yield was almost independent of the type of alcohol and exhibited a dramatic increase of 5--6 times with alcohol concentration.

The fluorescence decay times and corresponding weighting coefficients are given in Tables~\ref{tab:params_NADH} and \ref{tab:params_FAD} for NADH and FAD in water-alcohol solutions, respectively. The data in Table~\ref{tab:params_NADH} were obtained in our recent publication~\cite{Gorbunova2021SPIE} and the data in Table~\ref{tab:params_FAD} were obtained in this paper experiments. As can be seen in Tables~\ref{tab:params_NADH} and \ref{tab:params_FAD} the fluorescence decay in  both molecules was double exponential with decay times $\tau_1$ and $\tau_2$.

\begin{table}[h]
\caption{Fluorescence decay times $\tau_i$ in NADH determined upon two-photon excitation at 720~nm~\cite{Gorbunova2021SPIE}.}
\label{tab:params_NADH}
\begin{center}
\begin{tabular}{|c|c|c|}
\hline
\rule[-1ex]{0pt}{3.5ex} Solvent & $\tau_1$, ns ($a_1$) & $\tau_2$, ns ($a_2$)\\
\hline
\rule[-1ex]{0pt}{3.5ex} H$_2$O & 0.26 (0.74) & 0.62 (0.26)  \\
\hline
\rule[-1ex]{0pt}{3.5ex} MeOH 98\% & 0.30 (0.41) & 0.72 (0.59)  \\
\hline
\rule[-1ex]{0pt}{3.5ex} EtOH  98\% & 0.38 (0.33)  & 0.93 (0.67) \\
\hline
\rule[-1ex]{0pt}{3.5ex} PG  98\% & 0.56 (0.44) & 1.23 (0.56) \\
\hline
\end{tabular}
\end{center}
\end{table}

Experimental data in  Table~\ref{tab:params_NADH} for aqueous solution are in agreement with our recent results~\cite{Vasyutinskii2017,Gorbunova20c} and with those those reported earlier in
NADH under one-\cite{Couprie1994,Ladokhin1995} and two-photon~\cite{Blacker2019} excitation within experimental error bars. Also, both decay times shown in Table~\ref{tab:params_NADH} in pure methanol agree roughly with those reported earlier~\cite{Ladokhin1995,Krishnamoorthy1987} where a three-exponential model was used.

\begin{table}[h]
\caption{Fluorescence decay times $\tau_i$ in FAD determined upon  excitation at 355 nm and 450 nm.}
\label{tab:params_FAD}
\begin{center}
\begin{tabular}{|c|c|c|c|c|}
\hline
& \multicolumn{2}{c|}{ 355 nm } & \multicolumn{2}{c|}{ 450 nm}\\
\hline
\rule[-1ex]{0pt}{3.5ex} Solvent & $\tau_1$, ns ($a_1$) & $\tau_2$, ns ($a_2$) & $\tau_1$, ns ($a_1$)  & $\tau_2$, ns ($a_2$) \\
\hline
\rule[-1ex]{0pt}{3.5ex} H$_2$O & 2.17 (0.68) & 4.46  (0.32) & 2.40  (0.69) & 4.52 (0.31) \\
\hline
\rule[-1ex]{0pt}{3.5ex} MeOH 80 \% & 2.22 (0.32)  & 4.05 (0.68)  & 3.23 (0.61) & 4.58 (0.39) \\
\hline
\rule[-1ex]{0pt}{3.5ex} EtOH 80 \% & 2.51  (0.11) & 4.13  (0.68) & 2.57  (0.13) & 4.51  (0.87)\\
\hline
\rule[-1ex]{0pt}{3.5ex} PG 80 \% & 2.20 (0.11) & 4.97 (0.89)  & 3.01  (0.15) & 5.30 (0.75)\\
\hline
\end{tabular}
\end{center}
\end{table}

The existence of two excited state lifetimes in FAD obtained in this paper is in agreement with the results reported by Sengupta et al.~\cite{Sengupta2011} and Esposito et al.~\cite{Esposito2019}, however contradicts to the conclusions given by Nakabayashi et al.~\cite{Nakabatashi2010}, where the existence of four lifetimes have been reported under similar experimental conditions.  The decay time $\tau_1$ and $\tau_2$ values  in Table~\ref{tab:params_FAD} obtained in aqueous solution are in perfect agreement with those reported earlier~\cite{Sengupta2011,Esposito2019,Krasnopevtseva2020}.

\section{Discussion}
\label{sec:Discussion}

\subsection{A model of the excited state relaxation dynamics}
{We consider a simplified model taking into consideration radiative excitation and fluorescence channels as well as nonradiative excited state relaxation pathways. A generalized scheme of optical excitation in NADH and FAD via the first excited state $S_1$ and following relaxation pathways is shown in Fig.~\ref{scheme:PES}, where $R$ is the reaction coordinate and $S_0$ and $S_1$ are the PESs of the electronic ground and the first excited states, respectively.

\begin{figure} [ht]
   \begin{center}
   \begin{tabular}{c}
   \includegraphics[height=8cm]{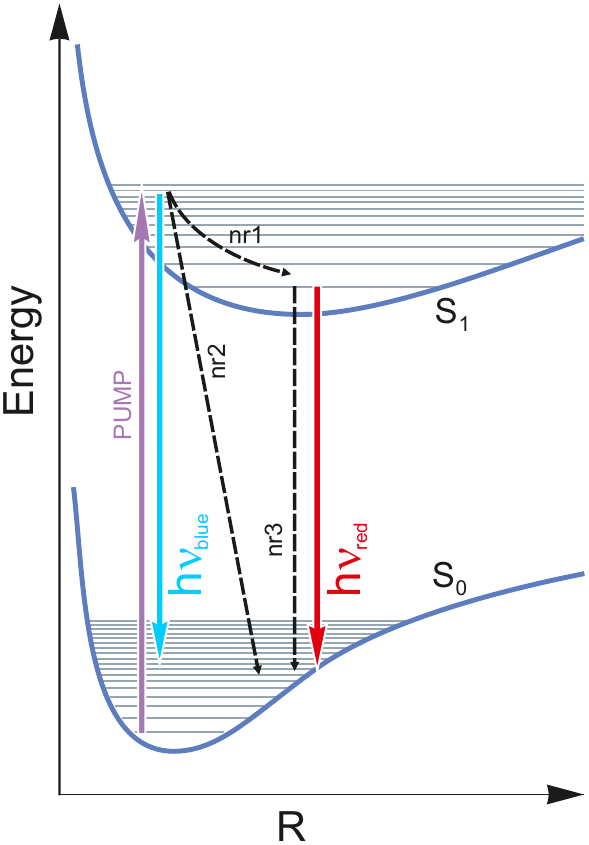}
   \end{tabular}
   \end{center}
   \caption{A generalized scheme of excitation and relaxation pathways in FAD and HADH.}
    \label{scheme:PES}
\end{figure}

According to Fig.~\ref{scheme:PES} an ultrashort laser pulse resonantly promotes ground state molecules to the highly excited vibrational energy levels of the first electronic excited state and then two fast picosecond nonradiative relaxation processes occur shown with black dashed lines with arrows. One of them is the vibrational relaxation (VR) within the excited state manifold that results in the population of the lowest vibrational levels of the state $S_1$.  This relaxation can be associated with the relaxation decay time of about $\tau_v \simeq$ 1~ps that was observed by several groups in NADH~\cite{Knutson2019,Knutson2020,Peon2020} and FAD~\cite{Radoszkowicz2011,Nakabatashi2010,vandenberg2002,Chosrowjan2003,Kondo2006,Kao2008}. This fast vibrational relaxation is schematically shown in Fig.~\ref{scheme:PES} with a black dashed line with arrow  denoted $nr1$.


The model suggested in this paper is based on the quantum-mechanical expressions describing transient absorption signals and fluorescence decay after one-, or two-photon excitation of a polyatomic molecule by an ultrafast laser pulse reported recently by Gorbunova et al.~\cite{Gorbunova20b} According to ref.~\cite{Gorbunova20b}, the contribution from the fast vibrational relaxation $nr1$ to the fluorescence decay reflects fast rearrangement of the molecular nuclear configuration in the molecular excited state after ultrashort laser excitation and depends on both internal molecular interactions and interactions with solute molecules. The corresponding contribution to the isotropic fluorescence decay can be approximated in the form~\cite{Gorbunova20b}:
\begin{eqnarray}
\label{eq:overlap4}
R^{fl}_0(t) \approx A^{iso}_1+(A^{iso}_0-A^{iso}_1)e^{-{t/\tau_{v}}},
    \end{eqnarray}
where time-independent terms $A^{iso}_0$ and $A^{iso}_1$  specify the values of the term $R^{fl}_0(t)$ at the nuclear
configurations that refer to the initial molecular state $S_0$ and to state $S_1$, respectively: $A^{iso}_0=R^{pr}_0(t)|_{S_0}$, $A^{iso}_1=R^{pr}_0(t)|_{S_1}$, and the term $\tau_{v}$ is the vibrational relaxation time.

\subsection{Decay associated spectra (DAS)}
The term  $A^{iso}_0$ is proportional to the fluorescence intensity from the excitation point (blue arrow in  Fig.~\ref{scheme:PES}) and the term $A^{iso}_1$  is proportional to the excitation intensity from the excited state equilibrium point (red arrow in  Fig.~\ref{scheme:PES}). Depending on the conditions $A^{iso}_0 > A^{iso}_1$, or $A^{iso}_0 < A^{iso}_1$, eq.~(\ref{eq:overlap4}) describes positive, or negative DAS.
 Also, the terms $A^{iso}_0$ and $A^{iso}_1$  are proportional to the corresponding Frank-Condon factors of the excited (\emph{e}) and ground (\emph{g}) electronic states: $A^{iso}_0\sim \langle v_e|v_g \rangle^2|_{t=0}$, $A^{iso}_1\sim \langle v_e|v_g \rangle^2|_{t\rightarrow\infty}$, where $v_g$ and $v_e$ are vibrational wave functions.

 The scheme illustrating the fluorescence decay during the fast vibrational relaxation $nr1$ in given in Fig.~\ref{scheme:PES}. As can be seen in Fig.~\ref{scheme:PES} Frank-Condon integrals $\langle v_e|v_g \rangle$ depend significantly on the radiative transition energy. In particular, if the photon energy exceeds the transition energy shown with a red arrow and denoted as $h\nu_{red}$, the fluorescence intensity from the condition $A^{iso}_0 > A^{iso}_1$ is fulfilled and therefore the blue wing of the fluorescence band is characterized by positive DAS. When the fluorescence photon energy decreases the factor $A^{iso}_1$ increases having its maximum at 0-0 transition, while the factor $A^{iso}_0$ decreases monotonously due to more and more unfavorable relationship between the vibrational wave functions $v_g$ and $v_e$.  Therefore the condition $A^{iso}_0 \leq A^{iso}_1$ can be fulfilled and the red wing of the fluorescence band can be characterized by near zero, or negative DAS. The conclusions made are in qualitative agreement with experimental observations~\cite{Radoszkowicz2011,Knutson2019,Knutson2020,Peon2020,Nakabatashi2010,vandenberg2002,Chosrowjan2003,Kondo2006,Kao2008}. The detailed description of DAS as function of the fluorescence photon energy depends on the actual PESs of the ground and excited states that as not known.   Note that the vibration relaxation occurs within the excited state and therefore does not contribute to the measured fluorescence quantum yield.

\subsection{Picosecond quenching dynamics}
Another type of picosecond relaxation occurs between the excited $S_1$ and the ground $S_0$ electronic states and results in fast fluorescence quenching that results in decrease of the observed quantum yield. This relaxation is shown in Fig.~\ref{scheme:PES} with a black dashed line with arrow denoted $nr2$. This relaxation can be associated with the picosecond decay times of about $\tau_d \simeq$ 26~ps observed in NADH~\cite{Knutson2019,Knutson2020} and of about $\tau_d \simeq$ 5-9~ps observed in FAD~\cite{Chosrowjan2003,Kao2008,Nakabatashi2010,Radoszkowicz2011}. In the case of NADH the routes of the quenching are still not understood in detail and in the case of FAD it is now well adopted~\cite{Islam2003,Li2008,Navarro2016,vandenberg2002,Radoszkowicz2010} that the mechanism of the fluorescence quenching occurs through an electron transfer reaction in the $\pi$-stacked conformation.


We suggest that irrespectively of the quenching mechanism the excited state population $N(t)$ in the course of QSSQ can be approximated by the expression:
  \begin{eqnarray}
\label{eq:N(t)}
N(t) \approx N_1 + N_2e^{-{t/\tau_{q}}},
    \end{eqnarray}
where concentrations $N_2$ and $N_1$ characterize the excited molecular conformations that are affected and not affected by QSSQ, respectively and $\tau_q$ is the quenching characteristic time.

In the case of FAD the concentration $N_2$ can be associated with the stacked conformations and the concentration $N_1$ can be associated with the unstacked conformations. Equation~(\ref{eq:N(t)}) describes the simplest case of interaction between two molecular states. In case of an exchange reaction between them, or if the number of interaction states is more that two, additional exponentials can appear in  eq.~(\ref{eq:N(t)}).

According to the model shown in Fig.~\ref{scheme:PES} when the fast picosecond relaxations is over and all remained conformations of excited molecules populate the lowest vibrational levels of the $S_1$ electronic state, much slow nanosecond radiative and nonradiative relaxations occur in the nanosecond and near subnanosecond time range.

\subsection{Time-resolved fluorescence decay in polyatomic molecules}
A general theoretical expression describing  fluorescence decay in polyatomic molecules excited by a short laser pulse can the approximated in the form:
\begin{eqnarray}
\label{eq:Iiso}
    I^{iso}_{fl}(t)&=&\frac{I_X + 2I_Y}{3}\approx{I_0}R^{fl}_0(t)N(t) \sum_l\,a_l\,e^{-t/\tau_l}.
\end{eqnarray}
where the term $R^{pr}_0(t)$ given in eq.~(\ref{eq:overlap4}) describes a contribution to the fluorescence signal from the fast vibrational relaxation, the term $N(t)$ given in eq.~(\ref{eq:N(t)}) describes QSSQ, $N(0)=N_1+N_2$, and the term under the sum with $l=1,2$ describes the fluorescence decay due to relatively slow nanosecond relaxation processes with decay times $\tau_1$ and $\tau_2$.

The sum in eq.~(\ref{eq:Iiso}) contains two exponentials with decay times $\tau_1$ and $\tau_2$ and weighting coefficients $a_1$ and $a_2$. The heterogeneity in the measured decay times in the nanosecond time domain in NADH and FAD in solutions and cells was well documented over a long period of time from the time-resolved fluorescence studies. Most of the authors agree that the heterogeneity in the measured fluorescence decay times in the nanosecond time domain in NADH and FAD contain  contributions from radiative and nonradiative decay channels. Particular routes of the nanosecond nonradiative relaxation in NADH and FAD are not known. These can be either ro-vibronic interactions in excited FAD molecules and interactions with surrounding solvent molecules, or rapid conformational fluctuations with intramolecular internal conversion to the ground electronic state. The former is shown in Fig.~\ref{scheme:PES} by a with a black dashed line with arrow denoted $nr3$ and the latter can occur in non-stacked conformations due to electron transfer reaction in the vicinity of CI. Anyway, each of the measured decay times: $\tau_1$ and $\tau_2$ can be presented as a combination of the radiative $\tau^{\,rad}$ and nonradiative $\tau^{\,nrad}$ decay times:
\begin{equation}
\label{eq:gammai}
 \frac{1}{\tau_l}=\frac{1}{\tau^{\,rad}}+\frac{1}{\tau^{\,nrad}_l},
\end{equation}
where the radiative decay time $\tau^{\,rad}$ in eq.~(\ref{eq:gammai}) is assumed to be the same for all excited conformations.

\subsection{Fluorescence quantum yield}
The fluorescence quantum yield $Q$ obtained by time-averaging of eq.~(\ref{eq:Iiso}) taking into account eqs.~(\ref{eq:overlap4}) and (\ref{eq:N(t)}) can be presented in the form:
\begin{equation}
\label{eq:Q}
Q=Q_0\frac{a_1\tau_1+a_2\tau_2}{\tau^{\,rad}},
\end{equation}}
with
\begin{equation}
\label{eq:Q0}
 Q_0=\frac{N_2}{N(0)},
\end{equation}
where $a_1\tau_1+a_2\tau_2$ is proportional to the total fluorescence intensity, $\tau^{\,rad}$ is proportional to the fluorescence intensity that could be measured in the absence of the nanosecond nonradiative relaxation, and  the condition $a_1+a_2=1$ was taken into account.

The partial quantum yield $0\leq Q_0\leq 1$ in eqs.~(\ref{eq:Q}) and (\ref{eq:Q0}) is a ratio between the number of molecules remaining in the bottom of the first excited electronic state $S_1$ after the picosecond dynamics is finished and the total concentration of the molecules excited by the laser pulse $N(0)=N_1+N_2$.

\subsection{Experimental data analysis}
\begin{figure} [ht]
   \begin{center}
   \begin{tabular}{c}
   \includegraphics[height=4.5cm]{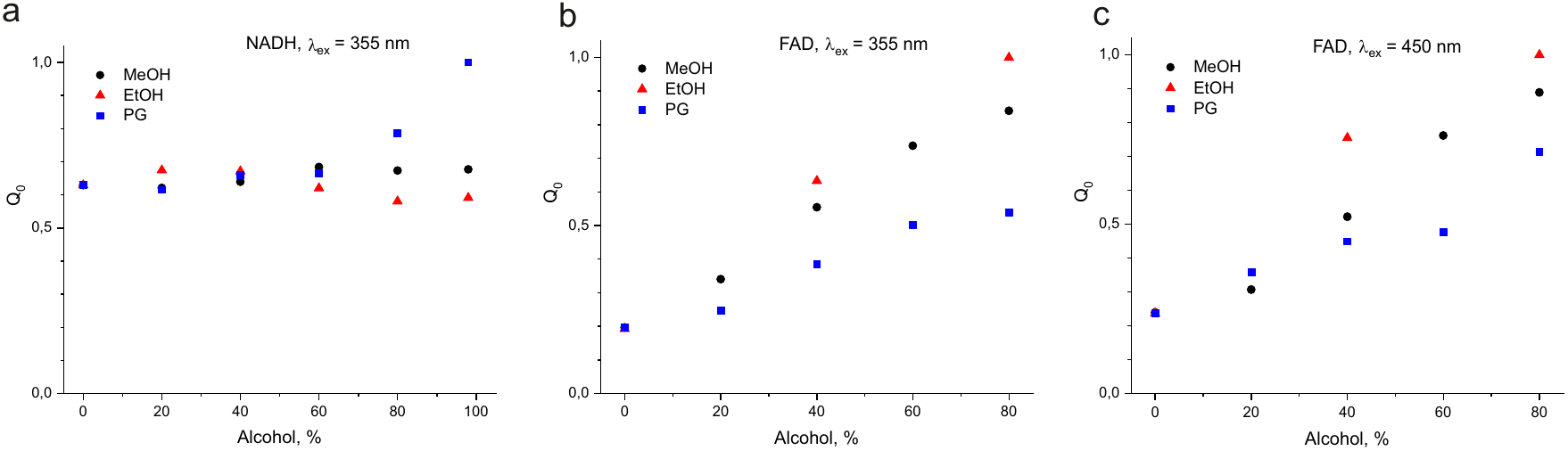}
   \end{tabular}
   \end{center}
   \caption{Partial quantum yield $Q_0$ in NADH and FAD as a function of the alcohole concentration in solution for three solutions: PG, MeOH, and EtOH. }
    \label{fig:Q0}
\end{figure}

The terms $Q$, $Q_0$, and $(a_1\tau_1+a_2\tau_2)$ in eq.~(\ref{eq:Q}) were assumed to be the function of alcohole concentration, while the radiative decay time $\tau^{\,rad}$ is assumed to be a constant. All three terms were determined from the experimental data presented in sec.~\ref{sec:experimental results}. Using the quantum yield experimental values in Figs.~\ref{fig:qy_FAD__alc} and \ref{fig:qy_NADH_alc} and the experimentally determined delay times $\tau_1$, $\tau_2$, and weighting coefficients $a_1$,  $a_2$  the ratio $Q_0/\tau^{\,rad}$ was determined from eq.~(\ref{eq:Q}) as a function of alcohole concentrations for all three alcohols used: MeOH, EtOH, and PG. The radiative time $\tau^{\,rad}$ and the partial quantum yield $Q_0$ were determined separately from the ratio $Q_0/\tau^{\,rad}$ using the the boundary condition $Q_0=1$ for the highest value of the ratio at the highest alcohol concentration. The boundary condition $Q_0=1$ was more sound in the case of FAD as it is well known that at high alcohole concentration fast picosecond fluorescence quenching in stacked FAD conformations give small, or negligible contribution to excited state dynamics~\cite{Nakabatashi2010,Radoszkowicz2010,Radoszkowicz2011}. In the case of NADH this boundary condition was less sound because although the dynamics of the fast picosecond fluorescence quenching in this molecule is known~\cite{Knutson2019,Knutson2020} its mechanisms are so far pure understood. Anyway, this approach gives correct behavior of $Q_0$ as a function of the alcohol concentration however can only result in some vertical shift in Fig.~\ref{fig:Q0}a below.

The obtained  partial quantum yields $Q_0$ for NADH and FAD are presented in Figs.~\ref{fig:Q0}a--c  as a function of the alcohol concentration for three solutions: PG, MeOH, and EtOH. The determined radiative decay times $\tau^{rad}$ are as follows. Fig.~\ref{fig:Q0}a: HADH excitation at 355~nm, $\tau^{rad}$=10.6~ns, Fig.~\ref{fig:Q0}b: FAD excitation at 450~nm, $\tau^{rad}$=21.9~ns, and Fig.~\ref{fig:Q0}c: FAD excitation at 355~nm, $\tau^{rad}$=29.6~ns. The former value can be compared with the theoretical value of the radiative decay time in NADH $\tau^{\,rad}=20.6$~ns reported by Scott et al.~\cite{Scott1970}. Therefore, in both NADH and FAD the estimated radiative decay times  $\tau^{rad}$ were found to be much longer than the measured fluorescence decay times $\tau_1$ and $\tau_2$  in Tables~\ref{tab:params_NADH} and \ref{tab:params_FAD}.


This result supports the well adopted conclusion~\cite{Scott1970} that nonradiative decay rate dominates over the radiative one, $(\tau_i^{\,nrad})^{-1} \gg (\tau^{\,rad})^{-1}$. Therefore the first term in the \emph{rhs} in eq.~(\ref{eq:gammai}) can be neglected compare with the second term meaning that the measured fluorescence decay times are mostly relate with nanosecond nonradiative decay.

Figure~\ref{fig:Q0} manifests a significant difference between the behavior of the reduced quantum yield $Q_0$ in NADH and FAD as a function of the alcohol concentration.  $Q_0$ in NADH in Fig.~\ref{fig:Q0}a is almost constant at all concentrations of MeOH and EtOH, and at moderate concentrations of PG. Only at the PG concentrations larger than 70\% the reduced quantum yield $Q_0$ increases but not much. Having in mind that the change of alcohol concentration in solution results in the change of conformation distribution, in particular toward the increase of the unfolded conformations number, the data in Fig.~\ref{fig:Q0}a suggest that fast picosecond quenching processes in the molecular excited states in NADH do to actually depend on the conformation distribution at least in metanol-water and ethanol-water solutions. This means that the picosecond decay in NADH does not likely occur through electron transfer in the stacking configuration but through other mechanisms. The same is likely true in PG solutions where a moderate increase of the reduced quantum yield at high PG concentrations may be due to slowdown of the fast quenching processes in high viscosity conditions.

Therefore, the linear growth of the measured quantum yield $Q$ in NADH in MeOH and EtOH solutions in Figs.~\ref{fig:qy_NADH_alc}b and c with alcohol concentration can be explained by relatively slow nanosecond nonradiative processes described by the term $a_1\tau_1 + a_2\tau_2$ in eq.~(\ref{eq:Q}). As reported recently by Gorbunova et al.~\cite{Gorbunova20c} and shown in Table~\ref{tab:params_NADH} the measured decay times $\tau_1$ and $\tau_2$ in NADH in water-MeOH solutions  only slightly depend on the methanol concentration, while the weighting coefficient $a_2$ that refers to the longer lifetime increased in about 1.6 times from pure aqueous to pure methanol conditions. It should be noted however, that we do not associate the longer time $\tau_2$ with the unfolded NADH conformation. As shown in ref.~\cite{Gorbunova20c} $a_2$ was practically constant and equal to $a_2$= 0.24 at MeOH concentrations below 70\% and rose up to 0.4 only at higher concentrations. At the same time, the contribution of the folded conformation $N_{fol}$ is known to decrease dramatically with MeOH concentration and approaches zero value at about 70-80\% MeOH~\cite{Oppenheimer1971,Hull2001,Gorbunova20c}.

Similar conclusions can be made about the behavior of the quantum yield $Q$ in NADH in water-ethanol solutions in Fig.~\ref{fig:qy_NADH_alc}c based on Fig.~\ref{fig:Q0}c, Table~\ref{tab:params_NADH}, and the dependence of the parameters $\tau_1$, $a_1$, $\tau_2$, $a_2$ on EtOH concentration reported in ref.~\cite{Gorbunova20a}.

As can be seen in Fig.~\ref{fig:Q0}a the reduced fluorescence quantum yield $Q_0$ of NADH in propylene glycol remains almost constant at small and middle PG concentrations smaller than 70\%  and increases by about 1.6 times at larger PG concentrations. According to eq.~(\ref{eq:Q}) sharp nonlinear increase of the measured $Q$ with PG concentration in Fig.~\ref{fig:qy_NADH_alc}a can be associated with decrease of both fast picosecond and relatively slow nanosecong quenching nonradiative rates.

As can be seen in Figs.~\ref{fig:Q0}b and c the behavior of the reduced quantum yield $Q_0$ in FAD as a function of the alcohol concentration differs from that in NADH. $Q_0$ rises almost linearly with alcohol concentration and increases at the concentration of 80\% by a factor of 3--4 under excitation at 450~nm and by a factor of 2.5--5 under excitation at 355~nm. This sharp increase of $Q_0$ manifests the decrease of fast picosecond quenching rate in FAD excited states. The mechanism of the picosecond fluorescence quenching in FAD is well established~\cite{Islam2003,Li2008,Navarro2016,vandenberg2002,Radoszkowicz2010}, it occurs through an electron transfer reaction in the $\pi$-stacked conformation. The probability of the electron transfer reaction depends dramatically on FAD conformation: it is very high in $\pi$-stacked conformations where the adenine and isoalloxazine rings are parallel in close proximity to each other and is practically zero in unfolded conformations~\cite{Chosrowjan2003,Kao2008,Nakabatashi2010,Radoszkowicz2011}. Therefore, the sharp increase of $Q_0$ with alcohol concentration  in Figs.~\ref{fig:Q0}b and c reflects the dramatic decrease of the contribution of $\pi$-stacked conformations in solution.

Having in mind eq.~(\ref{eq:Q}), the comparison of the fluorescence reduced quantum yield $Q_0$ in FAD in Fig.~\ref{fig:Q0}b,c with the measured total quantum yield $Q$ in Fig.~\ref{fig:qy_FAD__alc} suggests  that the fast picosecond dynamics gives the major contribution to the fluorescence signal intensity change with MeOH concentration, however there is a minor influence from the slow nanosecond dynamics.

\section{Conslusions}
The studies of the quantum yield and the dynamics of fluorescence decay in NADH and FAD excited at 355~nm and 450~nm in water- methanol, ethanol, and propylen glycole mixtures have been carried out. The global fit procedure was used for determination of the fluorescence parameters from experimental data.
A significant difference was observed in the behavior of NADH and FAD quantum yields in alcohol solution. In particular, the dependence of quantum yield in NADH on alcohol concentration was similar for monohydric alcohols (methanol and ethanol) and significantly different for polyhydric alcohol (propylene-glycol). In the case of FAD, the behavior of quantum yield was almost independent of the type of alcohol and exhibited a dramatic increase of 5--6 times with alcohol concentration. The contribution of the fast picosecond nonradiative decay in NADH to the quantum yield in water-MeOH and water EtOH mixtures was found to be almost independent of the alcohol concentration. The contribution of the fast picosecond nonradiative decay in NADH to the quantum yield in water-PG solutions was found to depend on PG concentration. The contribution of the fast picosecond quenching to the quantum yield in FAD was found to depend significantly on the alcohol concentration in all MeOH, EtOH.  The experimental results were analysed using a new model based on the quantum mechanical theory developed recently by the authors. The analysis of general theoretical expression describing  fluorescence decay in polyatomic molecules excited by a short laser pulse gave a new insight of the nature of DAS phenomenon. An expression for the fluorescence quantum yield that allowed to separate the contributions from the fast and slow nonratiative processes in molecular excited states was derived and used for clarification of the role of nanosecond and picosecond relaxation channels in NADH and FAD in various water-alcohol mixtures.  A sharp increase of the measured quantum yield with PG concentration was associated with decrease of both fast picosecond and relatively slow nanosecong quenching nonradiative rates.
As was shown the fast quenching gives a profound contribution to the rise of the measured quantum yield with alcohol concentration. The results obtained were compared with the results reported by other authors.

\section*{Acknowledgement}
The experiments were carried out under the support of Russian Science Foundation, grant No 18-53-34001. Calculations and models developed were performed under the financial support of BASIS Foundation, grant no. 19-1-1-13-1. The authors are grateful to the Ioffe Institute for providing the equipment used in experiments.

\bibliography{yield}

\end{document}